\documentclass[aps,prl,twocolumn,showpacs]{revtex4}
\usepackage{CJK}
\usepackage{bm}
\usepackage{graphicx}
\usepackage{color}
\usepackage{amsmath,amssymb,amsfonts}

\begin{document}

\title{Novel $\mathbb{Z}_2$-topological metals and semimetals}

\author{Y. X. Zhao}
\email[]{yuxinphy@hku.hk}

\author{Z. D. Wang}
\email[]{zwang@hku.hk}

\affiliation{Department of Physics and Center of Theoretical and Computational Physics, The University of Hong Kong, Pokfulam Road, Hong Kong, China}

\begin{abstract}

We report two theoretical discoveries for $\mathbb{Z}_2$-topological metals and semimetals. It is shown first that any dimensional
$\mathbb{Z}_2$ Fermi surface is topologically equivalent to a Fermi point. Then the famous conventional no-go theorem, which was merely proven  before for
$\mathbb{Z}$ Fermi points  in a periodic system without any discrete symmetry, is generalized to that the total topological charge is zero for all
cases.
 Most remarkably, we find and prove an unconventional strong no-go theorem: all $\mathbb{Z}_2$ Fermi points have the same topological charge
$\nu_{\mathbb{Z}_2} =1$  or $0$ for periodic systems.
 Moreover,  we also establish all six topological types of $\mathbb{Z}_2$ models for realistic physical dimensions.

\end{abstract}


\maketitle

\textit{Introduction}
 Topological semimetals,
such as  Weyl and some Dirac semimetals, have  recently attracted a huge research interest both theoretically and experimentally~\cite{Dirac-0,Z2-1,Dirac-1,Dirac-2,Dirac-3,Dirac-4,Dirac-5,Graphene,Xi,Hasan-1,Hasan-2,WSM-response,WSM1,WSM2,Zhao-Wang-WSM}.
As is known, topological metals and semimetals are essentially characterized  by the existence of Fermi surfaces (FSs) with nontrivial topological charges.
Several achievements have been made for topological classification and stability of FSs~\cite{Volovik-book,Horava,FS-classification,TI-FS}.
 In particular, anti-unitary symmetries, such as the time-reversal and/or particle symmetries (TRS and/or PHS), have recently been taken into account for classification of FSs~\cite{FS-classification,TI-FS}. Meanwhile, it has been noted that the TRS and/or PHS can lead to nontrivial $\mathbb{Z}_2$ topological charges, protecting inversion-invariant FSs against symmetry-preserving perturbations~\cite{FS-classification}.
 On the other hand, the implications of unitary symmetries, including the inversion and rotation, have also been explored~\cite{FP-1,FP-2,FP-3,FP-4}. These studies have deepened our understanding of symmetry-protected topological phases from gapped systems to gapless ones~\cite{TI-classification-1,TI-classification-2,TI-classification-3}. However, in-depth fundamental research on topological metals/semimetals consisting of $\mathbb{Z}_2$-FSs is still badly awaited, which may  not only  reveal a novel
 physics of $\mathbb{Z}_2$-topological metals and semimetals, but also pave the way for exploring exotic topological quantum matter.

 In this Letter, we first show that any FS protected by the above-captioned $\mathbb{Z}_2$ topological charge can continuously be  deformed to be a Fermi point with the symmetry being preserved \cite{note3}, and the survived Fermi point is still protected by a nontrivial topological charge~\cite{note-charge}. Based on this, we  elucidate that it is actually sufficient to consider only Fermi points in a $d$-dimensional ($d$D) system  with the spatial codimension $d_c=d-1$ (to be defined later), to exhaust all possible  inversion-invariant $\mathbb{Z}_2$ FSs. We then
generalize the well-known conventional no-go theorem, Eq.(\ref{No-go}), for $\mathbb{Z}$ Fermi points without any discrete symmetry in a given lattice model: these points appear in pairs having the opposite chiralities~\cite{No-go}, to that the sum of the topological charges of all $\mathbb{Z}$ or $\mathbb{Z}_2$ Fermi points vanishes.
For instance, in a Weyl semimetal, the net topological charge of a pair of left- and right-handed Weyl points is zero.
 As for a $d$D lattice model, there are $N=2^{d}$ inversion-invariant points for the $\mathbb{Z}_2$ Fermi points, and thus the above conventional no-go theorem may allow $2^{N-1}$ possible configurations of $\mathbb{Z}_2$ topological charges. In particular, we find and prove an unconventional strong no-go theorem, Eq.(\ref{Strong-no-go}), for such lattice models with $\mathbb{Z}_2$ Fermi points, which asserts that all of the points have the same topological charge $\nu_{\mathbb{Z}_2}$, leading to a real no-go status of the topological states with the only two possibilities.
Moreover,  we also construct various $\mathbb{Z}_2$ models in terms of  Dirac and Pauli matrices for illustration of physics, particularly including a simple lattice model consisting of 3D (or 2D) Dirac points that are protected by nontrivial $\mathbb{Z}_2$ topological charges with a spinful particle-hole symmetry.

\textit{Classification of Fermi Surfaces}
 We first review the general ideas underlying the classification of FSs~\cite{Volovik-book,FS-classification}. For a system $\mathcal{H}(k)$, choosing an $S^{d_c}$ in the gapped region of the momentum space, one can reveal the topological nature of the Berry connection restricted on the $S^{d_c}$. If its topology is nontrivial,  it is inevitable to meet gapless points when shrinking the $S^{d_c}$ to a point in any way. This means that the FS consisting of these gapless points is protected by the nontrivial topology, and this $d_c$ may be defined as the {\it spatial codimension} of the corresponding FS~\cite{note0}. The topological configurations are characterized by topological charges. In the complex Altland-Zirnbauer (AZ) classes \cite{AZClasses,AZClasses-2}, A and AIII, the topological charge is an integer $\nu_{\mathbb{Z}}\in\mathbb{Z}$. If the system has TRS and/or PHS, we can choose $S^{d_c}$ as the standard sphere centered at an inversion-invariant point in the $k$ space, and thus $\mathcal{H}(k)|_{S^{d_c}}$   has also the symmetries, noting that either TRS or PHS relates to an inversion operation changing $k$ to $-k$. The symmetries affect the topological properties of the Berry connection on $S^{d_c}$ in two aspects. First, for some cases in the presence of the symmetries, the topological charge $\nu_{\mathbb{Z}}$ is always trivial ($=0$). Second, due to the restriction of the symmetries, there are new topological configurations characterized by a symmetry-related topological charge $\nu_{\mathbb{Z}_2}\in\mathbb{Z}_2$, analogous to topological insulators with TRS. In shrinking the $S^{d_c}$ symmetrically to the inversion-invariant point, a nontrivial $\nu_{\mathbb{Z}_2}$ ensures topological robustness of gapless points, namely, the FS is topologically protected by the
$\mathbb{Z}_2$ charge.

\begin{figure}
	\includegraphics[scale=0.55]{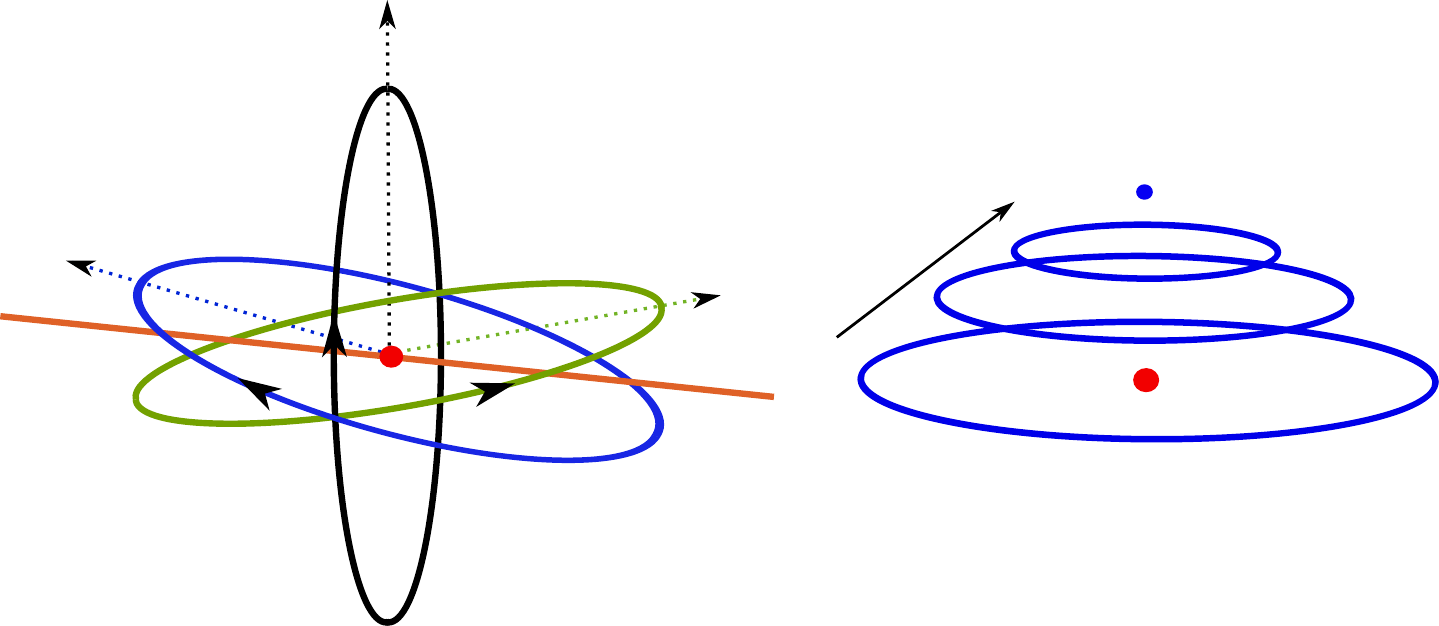}
	\caption{Opposite rotations of a circle and asymmetrically shrinking of a circle.\label{Rotations}}
\end{figure}

\begin{figure}
\includegraphics[scale=0.55]{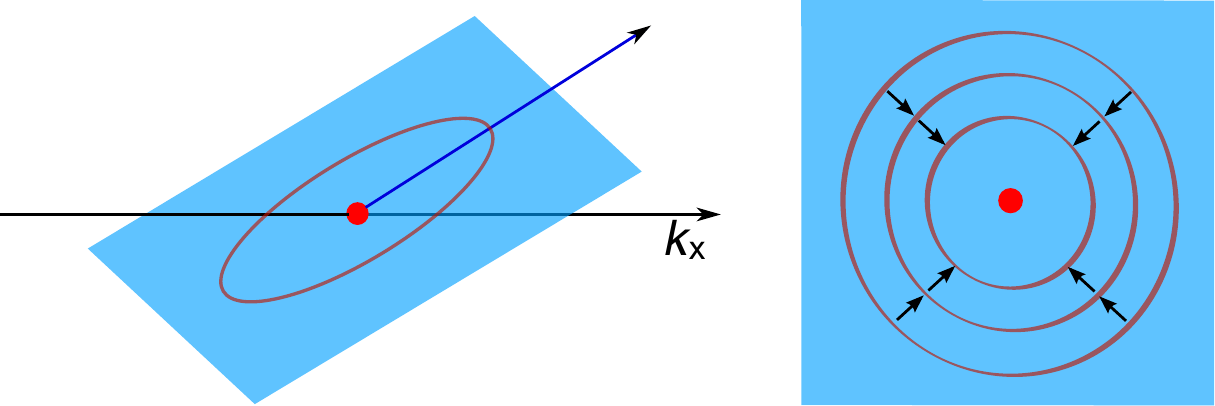}
\caption{Symmetric circles enclosing a $\mathbb{Z}_2$ Fermi point.\label{Circles}}
\end{figure}

\textit{Representative $\mathbb{Z}_2$ Fermi points}
For illustration of real physics, we address representative Fermi points of nontrivial $\mathbb{Z}_2$ topological charges in physical dimensions $d=1,2,3$ \cite{TI-FS}. As we will prove that all $\mathbb{Z}_2$ Fermi surfaces can be deformed to be Fermi points with unique codimension, this actually exhausts all possibilities. There are six nontrivial cases, which are conveniently divided into two groups. The three model Hamiltonians in the first group are built of $4\times 4$ Dirac matrices with two being 3-dimensional and one being 2-dimensional, which are given by $\mathcal{H}_{C}^{(2)}=\sum_{a=1}^3 k_a(\Gamma^a+\lambda \Gamma^5)$, $\mathcal{H}_{CII}^{(1),c}=\sum_{a=1}^3 k_a\Gamma^a$ and $\mathcal{H}_{CII}^{(2),c}=\sum_{a=1}^2 k_a\Gamma^a$, where $\sigma$ and $\tau$ are the two sets of Pauli matrices, the superscripts `$(1,2)$' denote $\mathbb{Z}_2^{(1,2)}$ and `$c$' indicates the presence of a chiral symmetry (CS), and the subscripts are the symmetry class names~\cite{FS-classification}. The Dirac matrices are defined as $\Gamma^a=\sigma^a\otimes\tau^1~(a=1,2,3)$, $\Gamma^4=1\otimes\tau^2$ and $\Gamma^5=1\otimes\tau^3$, and correspondingly  $\hat{T}=-i\sigma^2\otimes 1\hat{\kappa}$, $\hat{C}=-i\sigma^2\otimes\tau^3\hat{\kappa}$, and $S=\Gamma^5$ respectively for TRS, PHS, and CS, where $\hat{\kappa}$ is the complex conjugate operator. The second group is constructed by Pauli matrices $\tau_a$, including $\mathcal{H}_{AII}^{(1)}=\sum_{a=x}^yk_a (\tau_a+\lambda_a \tau_z)$, $\mathcal{H}_{AII}^{(2)}=k_x(\tau_x+\lambda_1 \tau_y+\lambda_2 \tau_z)$ and $\mathcal{H}_{DIII}^{(1),c}=k_x (\tau_x +\lambda_3 \tau_y)$,
with $\hat{T}=-i\tau_y\hat{\kappa}$, $\hat{C}=\tau_x\hat{\kappa}$ and $S=\tau_z$, where $|\lambda_x|+|\lambda_y| \neq 0$ and $\lambda_1 \lambda_2 \neq 0$.

\textit{$\mathbb{Z}_2$ Fermi surfaces}
Remarkably, the $\mathbb{Z}_2$ topological charges possess a unique feature that makes them distinct from  $\mathbb{Z}$ ones. Namely in contrast to $\mathbb{Z}$, every such $\mathbb{Z}_2$ Fermi surfaces can be reduced as a Fermi point by symmetry-preserving perturbations. As an example, considering a Fermi line of the $k_x$ axis for the Hamiltonian $\mathcal{H}=k_y(\tau_y+\lambda\tau_x)+k_z\tau_z$ in a 3D $k$ space, which is protected by a nontrivial $\nu_{\mathbb{Z}_2}^{(1)}$ in the class AII, a TRS preserving perturbation $\mathcal{H}'=\lambda' k_x\tau_x$  can distort the Fermi line into a Fermi point with the unit topological charge. To prove the general result but without loss of generality, let us consider a Fermi line in a three-dimensional $k$ space as shown in the left part of Fig.(\ref{Rotations}), where the Fermi line is the whole $k_x$ axis denoted by the orange line  with the red point being an inversion-invariant point. We choose a circle with a given orientation on the $k_y$-$k_z$ plane and rotate it with respect to $k_y$ axis respectively for $\pm(\pi/2-\epsilon)$ with $\epsilon$ being an infinitesimal constant,  resulting in the green and blue circles. The rotations are continuous transformations with the anti-unitary symmetries being preserved, which implies that the three circles have the same topological charge. It is observed that the green and blue circles tend to be of opposite orientations. If we  consider a $\mathbb{Z}$ topological charge $\nu_{\mathbb{Z}}$, reversing the orientation of one circle leads to $\nu_{\mathbb{Z}}\rightarrow-\nu_{\mathbb{Z}}$. Thus nonzero $\nu_{\mathbb{Z}}$ causes that there are topological obstructions to go across the Fermi line, and the Fermi line is well protected by $\nu_{\mathbb{Z}}$. In contrast, if it is the $\mathbb{Z}_2$ topological charge $\nu_{\mathbb{Z}_2}$ under consideration, reversing the orientation does not change $\nu_{\mathbb{Z}_2}$, since $-1\equiv 1 \mod 2$. Accordingly, no topological obstruction prevents one to move one circle across the Fermi line, or equivalently the Fermi line can be gapped by symmetry-preserving perturbations. The topological essence underlying the example lies in a fact that $\mathbb{Z}_2$ topological charges are defined by extensions \cite{FS-classification,Xiao-Liang-PRB}. Thus for a $d_{FS}$-dimensional ($0<d_{FS}<d$ with $d$ as the  dimension of $k$ space)
inversion-invariant FS with a nontrivial $\nu_{\mathbb{Z}_2}$ on the $S^{d_c}$~\cite{Note-FS}, one can always continuously extend the $S^{d_c}$ to a higher-dimensional sphere by crossing the FS away from the inversion-invariant plane of the $S^{d_c}$, namely, all gapless points on the FS, except the inversion-invariant point, can be gapped out without breaking the symmetries of the system.  Consequently,  we have an important conclusion that {\it the $\mathbb{Z}_2$ topological charges can only protect Fermi points}, while $\mathbb{Z}$ topological charges may protect  FSs with any dimension.

However, a new question arises from the above discussions, namely, a Fermi point in a $d$D $k$ space might have more than one spatial codimensions for $\mathbb{Z}_2$ topological charges. Explicitly, for possibly different $d_c$'s in the range of $0\le d_c<d-1$, if any $S^{d_c}$   has a nontrivial $\nu_{\mathbb{Z}_2}$, the Fermi point is protected topologically. For instance, in the above Fermi line case, the left gapless point at the inversion-invariant point is indeed protected by a nontrivial $\nu_{\mathbb{Z}_2}$, because, as shown in Fig.(\ref{Circles}), shrinking any inversion-invariant circle symmetrically will eventually meet gapless points as topological defects in the $k$ space. Nevertheless we can show that, due to the hierarchy among topological charges, it is sufficient to consider only the spatial codimension $d_c=d-1$ for $\mathbb{Z}_2$ Fermi points~\cite{note0}, with technical details being presented in the Supplemental Material~\cite{Supp}.

\textit{Lattice Models}
We now turn to consider lattice models with multiple topological Fermi points in the Brillouin zone(BZ). As we will see, the local topological charge for each Fermi point may have global implications in the whole BZ. To construct all relevant lattice models in physical dimensions, we merely need to directly replace $k_a$ by $\sin k_a$ in the representative continuous Dirac models, so that the coarse-grain expansion around each inversion-invariant point in the resulting lattice model is the corresponding $\mathbb{Z}_2$ Fermi point. We will see that this is actually a complete construction for lattice models with $\mathbb{Z}_2$ Fermi points, which is ensured by our unconventional strong no-go theorem, Eq.(\ref{Strong-no-go}). We now work out an interesting 3D model of  Dirac points in detail to see its physical meaning clearly, which reads $\mathcal{H}_D(k)=\sum_{a=1}^3 \sin k_a \Gamma^a$, or in the real space,
\begin{equation}
H_D=\sum_{\mathbf{j}}\sum_{a=1}^3it~\psi^\dagger_{\mathbf{j}+\mathbf{e}_a}\sigma^a\otimes\tau_1\psi_{\mathbf{j}}+h.c. , \label{3D-Lattice}
\end{equation}
where $\psi_{\mathbf{j}}=(c^A_{\mathbf{j}\uparrow},~c^A_{\mathbf{j}\downarrow},~c^B_{\mathbf{j}\uparrow},~c^B_{\mathbf{j}\downarrow})^T$, and $t$ denotes the hopping coefficient. It is found that the TRS $\hat{T}=-i\sigma^2\hat{\kappa}$ is just the standard spinful one and the CS $\Gamma^5=\tau^3$ represents the sublattice symmetry between sublattices $A$ and $B$, which may be regarded as fundamental symmetries of the lattice system, while the spinful PHS denotes a combined symmetry $\hat{C}=\Gamma^5\hat{T}=-i\sigma^2\otimes\tau^3\hat{\kappa}$.   Perturbations preserving the combined PHS, such as $\lambda \sin k\Gamma^5$, may be added to break the TRS and CS, meanwhile the topological protection of Dirac points still survives, but with the topological class being loosened from $\mathbb{Z}^{(1)}_2$ in the class CII with CS to $\mathbb{Z}^{(2)}_2$ in the class C with only the PHS. In addition, a relevant 2D lattice model for $\mathbb{Z}_2^{(2)}$ in the class CII may readily be obtained by a dimension reduction of Eq.(\ref{3D-Lattice}), namely, $a$ is  summed over $1$ and $2$.  It is noted that the $\mathbb{Z}_2$ topological charges for Fermi points addressed here are associated with the anti-unitary TRS or PHS, which are essentially different from those protected by the unitary spatial symmetries studied recently~\cite{Dirac-0,Z2-1}, leading to distinct implications to the stability against disorders~\cite{note2}.

\textit{No-go theorem} At this stage, we look at the
conventional no-go theorem for chiral Fermi points in a lattice model in terms of topological charges. A chiral Fermi point is just a Fermi point with a $\mathbb{Z}$ topological charge $\nu_{\mathbb{Z}}=\pm 1$ in the class A, where $`\pm'$ corresponds to the left- or right-handed chirality. In a $(2n+1)$D lattice model, the first Brillouin zone (BZ) is topologically a $(2n+1)$D torus $T^{2n+1}$.
 We consider  a finite number of Fermi points distributing on the $T^{2n+1}$. For the $j$th Fermi point, we can choose a $2n$D sphere $S_j^{2n}$ in its neighborhood enclosing only the Fermi point, and  evaluate its topological charge $\nu_j\in \mathbb{Z}$ whenever an orientation for the $S_j^{2n}$ is given. Since the BZ $T^{2n+1}$ is an orientable manifold, the orientation can be defined globally, i.e.,  for a given $j_0$ and viewed from all of the other Fermi points, the $S^{2n}_{j_0}$ encloses them with an opposite orientation. Since the $S^{2n}_{j_0}$ can be continuously deformed to be the collection of the all  other $S^{2n}_j$'s with $j\ne j_0$, it is found that $\nu_{j_0}=-\sum_{j\ne j_0}\nu_j$, or equivalently
\begin{equation}
\sum_j \nu_j=0, \label{No-go}
\end{equation}
recalling that reversing the orientation reverses the topological charge. The surgeries that make a torus into a sphere are presented in Supplemental Material~\cite{Supp}.


A Fermi point with topological charge $|\nu|>1$ can always be continuously deformed to be a collection of unit Fermi points with the total topological charge being $\nu$\cite{Volovik-book}. Since a unit Fermi point cannot be divided further, it is more stable than the ones with $|\nu|>1$ \cite{note1}. Thus Eq.(\ref{No-go}) just implies that a lattice model has stably pairs of left- and right-handed Fermi points, which is just the conventional
no-go theorem~\cite{No-go}. Eq.(\ref{No-go}) also holds for the class AIII with a chiral symmetry in even spatial dimensions, where a Fermi point may have a $\mathbb{Z}$ topological charge that depends on the chiral symmetry.  Consequently,  Eq.(\ref{No-go}) is a more appropriate way to state the no-go theorem for lattice models in the complex classes A and AIII \cite{note1}. In addition, as for the $\mathbb{Z}$ Fermi points of the other eight real symmetry classes~\cite{FS-classification}, it is also noted that Eq.(\ref{No-go}) is still valid.


\begin{figure}
\includegraphics[scale=0.6]{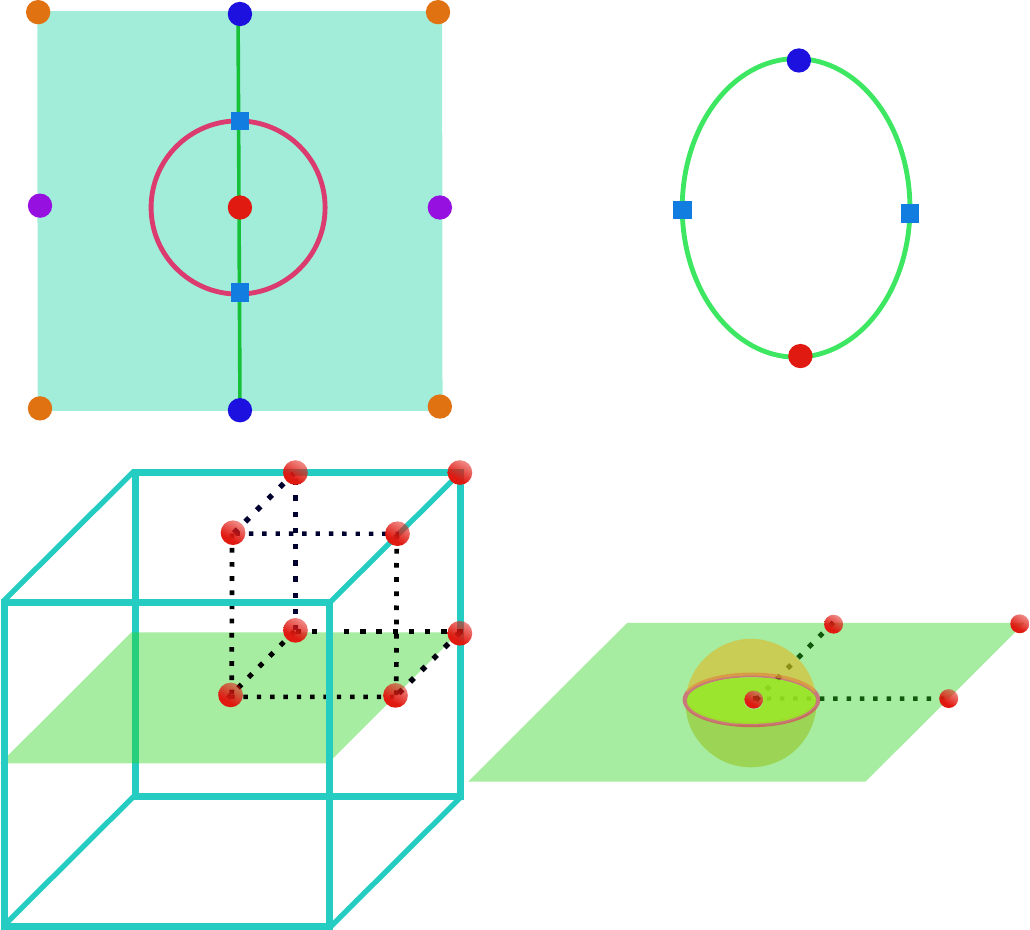}
\caption{Sphere reductions in inversion-invariant subspaces. Upper and lower parts correspond to 2D and 3D cases, respectively. \label{Sphere-reduction}}
\end{figure}

\textit{Unconventional strong no-go theorem} We now proceed to consider $\mathbb{Z}_2$ Fermi points at inversion-invariant points in a BZ.  For a given invariant point in a $d$D BZ, we choose an inversion-invariant $S^{d-1}$ to enclose it in its neighborhood. Then the $S^{d-1}$ can also be continuously deformed in an inversion-invariant way to be a collection of inversion-invariant $S^{d-1}$s, where every one of the other invariant points is enclosed by one of $S^{d-1}$s. An example of a 2D square BZ is shown in \cite{Supp}. Thus Eq.(\ref{No-go}) holds also for $\mathbb{Z}_2$ Fermi points. For $\mathbb{Z}_2$ Fermi points with the charge $-1\equiv 1\mod 2$, Eq.(\ref{No-go}) merely indicates that there are even number of nontrivial $\mathbb{Z}_2$ Fermi points.  In general,  there are $2^{N-1}=\sum_{n=0}^{N/2}C_N^{2n}$ possibilities for the total $N=2^d$ inversion-invariant points, where $C^{2n}_{N}$ is the number of distinct ways to choose $2n$ ones among $N$ elements. For example, in the $3$D model, Eq.(\ref{3D-Lattice}), with eight nontrivial $\mathbb{Z}_2$ Fermi points, one might expect that by adding certain symmetry-preserving terms, a pair of Fermi points would be gapped while the others still exist, which is allowed by Eq.(\ref{No-go}). However, it turns out that all of such trials should fail. As we will show later, there actually are only two possibilities for $\mathbb{Z}_2$ Fermi points, namely, these topological charges $\nu_{\mathbb{Z}_2}$ at inversion-invariant points are all nontrivial or trivial
\begin{equation}
\nu_{\mathbb{Z}_2,1}=\cdots=\nu_{\mathbb{Z}_2,j}=\cdots=\nu_{\mathbb{Z}_2,N}=0~~\mathrm{or}~~1, \label{Strong-no-go}
\end{equation}
where $\nu_{\mathbb{Z}_2,j}$ is the topological charge of the $j$th inversion-invariant point. This strong no-go theorem, Eq.(\ref{Strong-no-go}), for $\mathbb{Z}_2$ Fermi points serves as one of our main results.

The essential reason lies in the hierarchy of topological charges, which has been used  for demonstration of the sufficiency of the afore-mentioned spatial codimension \cite{Supp}. We shall first prove Eq.(\ref{Strong-no-go}) for $\mathbb{Z}_2^{(1)}$ topological charges, and then deduce it readily for the $\mathbb{Z}_2^{(2)}$ ones.  For clarity, we first work out a simple example of a 2D square lattice as shown in the upper part of Fig.(\ref{Sphere-reduction}), where each color refers to an independent inversion-invariant point. We consider the $\nu_{\mathbb{Z}_2,r}^{(1)}$ of the red point at the center, which is enclosed by a standard circle $S^{1}$. The green line through the point is an inversion-invariant line that intersects the $S^{1}$ on $S^{0}$ consisting of two blue square points. It is observed that the $\mathbb{Z}_2^{(2)}$ topological charge $\nu_{\mathbb{Z}_2,r}^{(2)}$ on the $S^{0}$ is equal to $\nu_{\mathbb{Z}_2,r}^{(1)}$, $\nu_{\mathbb{Z}_2,r}^{(2)}=\nu_{\mathbb{Z}_2,r}^{(1)}$ due to the hierarchy of $\mathbb{Z}_2$ topological charges, because $\mathcal{H}(k)|_{S^{1}}$ is a symmetric continuation of $\mathcal{H}(k)|_{S^{0}}$. The green line is actually a circle as shown in the upper-right part of Fig.(\ref{Sphere-reduction}), on which there is also the other inversion-invariant point in dark blue. The dark blue point is also enclosed by the $S^{0}$, implying that $\nu_{\mathbb{Z}_2,b}^{(2)}=\nu_{\mathbb{Z}_2,r}^{(2)}$, where $\nu_{\mathbb{Z}_2,b}^{(2)}$ is the $\mathbb{Z}_2^{(2)}$ topological charge of the point. Then we find $\nu_{\mathbb{Z}_2,b}^{(1)}=\nu_{\mathbb{Z}_2,b}^{(2)}$, through continuously extending the $S^{0}$ to an inversion-invariant circle enclosing only the dark-blue point in the 2D BZ. Thus $\nu_{\mathbb{Z}_2,b}^{(1)}+\nu_{\mathbb{Z}_2,r}^{(1)}=0$ on the inversion-invariant green line. Since there are many other ways to choose the inversion-invariant line, where each gives such an equation, we can have three independent equations of $\mathbb{Z}_2^{(1)}$ topological charges to prove Eq.(\ref{Strong-no-go}). For example, apart from the green line, we can choose an inversion-invariant horizontal line and a diagonal line.

The 2D example can be generalized to any dimensions readily. In any $(d-1)$D inversion-invariant sub-BZ $SBZ^{a}$ of a $d$D BZ with $a$ being the sub-BZ index, an $S^{(d-1)}$ enclosing a $\mathbb{Z}_2^{(1)}$ Fermi point is reduced to be a $S^{(d-2)}$ that has a $\mathbb{Z}_2^{(2)}$ topological charge $\nu_{\mathbb{Z}_2}^{(2)}=\nu_{\mathbb{Z}_2}^{(1)}$. Since restricted on the $SBZ^{a}$, the $S^{(d-2)}$ can continuously be  deformed to be the other $S^{(d-2)}$s enclosing the rest inversion-invariant points on the sub-BZ, leading to $\sum_{j\in SBZ^{a} }\nu_{\mathbb{Z}_2,j}^{(2)}=0$, namely a $\mathbb{Z}_2^{(2)}$ no-go theorem on $SBZ^{a}$.  Extending continuously each $S^{d-2}$ into $S^{d-1}$ in the whole BZ, we obtain a $\mathbb{Z}_2^{(1)}$ no-go theorem on $SBZ^{a}$,
\begin{equation}
\sum_{j\in SBZ^{a} }\nu_{\mathbb{Z}_2,j}^{(1)}=0, \quad \mathrm{for}~~\mathrm{all}~~SBZ^{a}. \label{Sub-no-go}
\end{equation}
A 3D cube example is shown in the lower part of Fig.(\ref{Sphere-reduction}) to illustrate the general result of
Eq.(\ref{Sub-no-go}). Since there always are $N-1$ independent choices of the inversion-invariant sub-BZ contributing $N-1$ independent equations of Eq.(\ref{Sub-no-go}), we have proved the strong no-go theorem of Eq.(\ref{Strong-no-go}) for $\mathbb{Z}_2^{(1)}$ topological charges to be the only two solutions of Eq.(\ref{Sub-no-go}). The $\mathbb{Z}_2^{(2)}$ part of Eq.(\ref{Strong-no-go}) can also be deduced by noticing that a $\mathbb{Z}_2^{(2)}$ system can alway be regarded as a subsystem of a $\mathbb{Z}_2^{(1)}$ system~\cite{Supp}.

In summary, we have not only revealed unambiguously that $\mathbb{Z}_2$ topological charges can only protect Fermi points, but also proved  the unconventional strong no-go theorem for  $\mathbb{Z}_2$ Fermi points.
Moreover, relevant lattice modes have also been established.

\begin{acknowledgments}
{\it Acknowledgments} We thank A. P. Schnyder and X. Dai for helpful discussions. The work  was supported by the GRF (Grant Nos.  HKU173051/14P and HKU173055/15P) and the CRF (HKU8/11G) of Hong Kong.
\end{acknowledgments}

\section{Supplemental Material}

\subsection{The Hierarchy among topological charges }
\begin{figure}
	\includegraphics[scale=0.55]{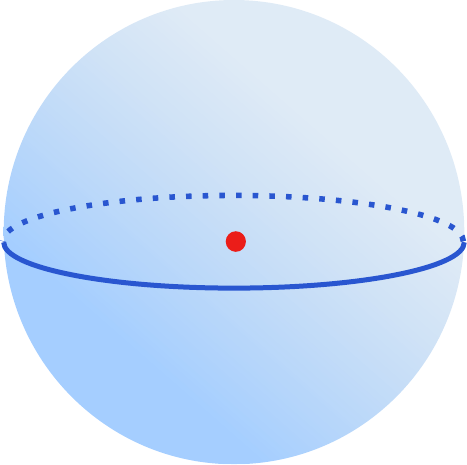}
	\caption{A circle as a symmetric extension of a sphere.\label{Extension}}
\end{figure}

We recall that, among our complete set of topological charges, a subset of a $\mathbb{Z}$ and two $\mathbb{Z}_2$ topological charges form a topological family $\nu_{\mathbb{Z}}$, $\nu_{\mathbb{Z}_2}^{(1)}$ and $\nu_{\mathbb{Z}_2}^{(2)}$ for the spatial codimensions in decreasing orders, $d_c$, $d_c-1$, $d_c-2$, respectively ~\cite{FS-classification,Xiao-Liang-PRB}. $\nu_{\mathbb{Z}_2}^{(1)}$ is defined as the parity of $\nu_{\mathbb{Z}}$ for a continuous symmetry-preserving extension of $\mathcal{H}|_{S^{d_c-1}}$ from $S^{d_c-1}$ to $S^{d_c}$, and analogously $\nu_{\mathbb{Z}_2}^{(2)}$ is defined as $\nu_{\mathbb{Z}}^{(1)}$ for a continuous symmetry-preserving extension of $\mathcal{H}|_{S^{d_c-2}}$ from $S^{d_c-2}$ to $S^{d_c-1}$, as illustrated in Fig.(\ref{Extension}). For topological reasons, such extensions can only be done up to two steps for well-defined topological invariants.

\subsection{Unique $d_c=d-1$ for $\mathbb{Z}_2$ Fermi points}
We now elucidate that it is sufficient to consider $d_c=d-1$ for $\mathbb{Z}_2$ Fermi points
from the inherent hierarchy of the topological charges~\cite{FS-classification,Xiao-Liang-PRB}. Without loss of generality, we address in detail a nontrivial case in physical dimensions. For a  $\mathbb{Z}_2$ Fermi point in the 3D $k$ space, it is enclosed inversion-invariantly either by a sphere $S^2$ or by a circle $S^1$. Noting that the circle can always be continuously deformed to be an equator of the sphere, $\mathcal{H}(k)|_{S^2}$ can be regarded as a continuous extension of $\mathcal{H}(k)|_{S^1}$, with all of the anti-unitary symmetries being preserved. Due to the hierarchy among topological charges, it is now clear that a nontrivial  $\nu_{\mathbb{Z}_2}^{(1)}$ on the circle implies that $\nu_{\mathbb{Z}}$ is an odd number on the sphere, and similarly a nontrivial  $\nu_{\mathbb{Z}}^{(2)}$ on the circle is equivalent to a nontrivial $\nu_{\mathbb{Z}}^{(1)}$ on the sphere. Since $\nu_{\mathbb{Z}}$ ($\nu_{\mathbb{Z}_2}^{(1)}$) contains more topological information than $\nu_{\mathbb{Z}_2}^{(1)}$ ($\nu_{\mathbb{Z}_2}^{(2)}$) does, it is sufficient to deal with only the spatial codimension $d_c=2$ in this case.  For a Fermi point in $d$D $k$ space with $d$ being large enough, it is clear from the above example that  we need not to consider $d_c=d-2$ ($d_c=d-2,d-3$) for $\mathbb{Z}_2^{(1)}$($\mathbb{Z}_2^{(2)}$). But for $d_c<d-2$ ($d_c<d-3$), $\nu_{\mathbb{Z}_2}^{(1)}$($\nu_{\mathbb{Z}_2}^{(2)}$) always vanishes, because $S^{d_c}$ can be symmetrically extended in the $k$ space to be $S^{d_c+1}$($S^{d_c+2}$) with the corresponding $\mathbb{Z}$ topological charge being zero because of $d_c+1<d-1$ ($d_c+2<d-1$). To conclude,  it is sufficient to consider $d_c=d-1$ for Fermi points.

\subsection{Surgeries from a torus to a sphere}
\begin{figure}
	\includegraphics[scale=1.1]{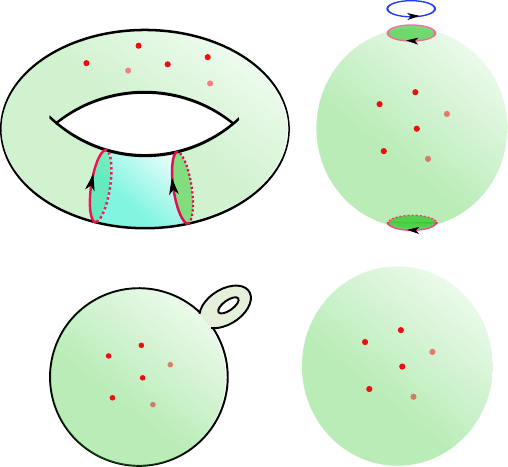}
	\caption{The deformation of a torus to a sphere.\label{Torus}}
\end{figure}

As a topologically equivalent picture of the torus, Fig.(\ref{Torus}) shows the surgeries making a torus to a sphere through cutting a segment without Fermi points. In accord with the global orientation of the sphere, the direction of the upper circle has to be reversed with respect to that of the lower one, cancelling  with each other, namely, the handle of the torus can always be cut off to become a sphere in the discussion of the total topological charge of all Fermi points.

\subsection{An Example: Symmetric deformation of a circle}
\begin{figure}
	\includegraphics[scale=0.6]{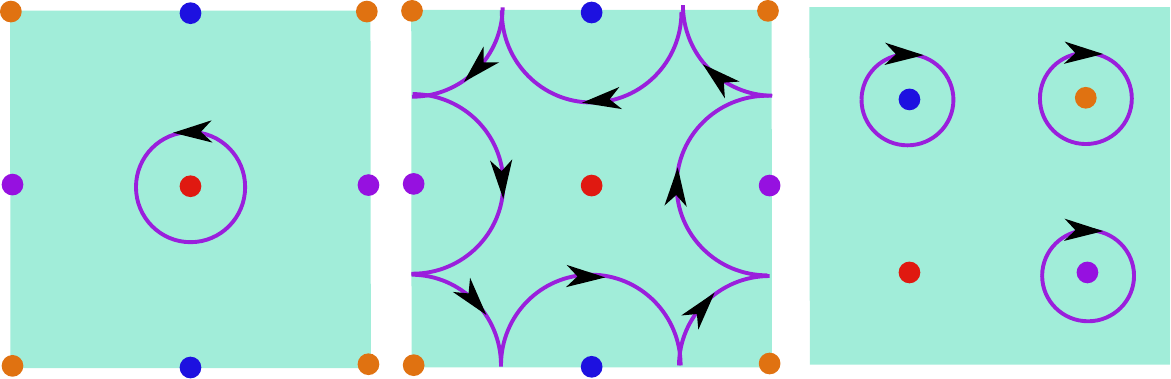}
	\caption{The symmetric deformation of a circle with three steps from left to right.\label{Circle-deform}}
\end{figure}

An inversion-invariant $S^{d_c}$ chosen to enclose a Fermi point in a Brillouin zone can always be continuously deformed in an inversion-invariant way to be a collection of spheres enclosing the other Fermi points with the opposite orientation. This is illustrated in Fig.(\ref{Circle-deform}) in a two-dimensional Brillouin zone.

\subsection{Proof of Eq.(3) for the $\mathbb{Z}_2^{(2)}$ part}

We here deduce the $\mathbb{Z}_2^{(2)}$ part of Eq.(3). For any $(d-1)$D Hamiltonian  with $\mathbb{Z}_2^{(2)}$ Fermi points at inversion-invariant point, we can always regard it as a $d$D Hamiltonian of $\mathbb{Z}_2^{(1)}$ Fermi points being restricted on an inversion-invariant sub-BZ, since the condition of Eq.(4) is satisfied. The strong no-go theorem of Eq.(3) for $\mathbb{Z}_2^{(1)}$ implies that  all Fermi points have the same $\mathbb{Z}_2^{(1)}$ topological charge in the extended $d$D model. Reducing back to the inversion-invariant sub-BZ of the original system, Eq.(3) has been deduced for the $\mathbb{Z}_2^{(2)}$ Fermi points. 


\begin{thebibliography}{References}
\bibitem{note3} A Fermi point here means an energy level-crossing point. The Fermi level is not necessarily located at the crossing point, except that is required by the chiral or particle-hole symmetry.
\bibitem{note-charge} Due to the hierarchy nature for topological charges, the topological protection cabability  is enhanced in the order $\mathbb{Z}_2^{(2)}\rightarrow\mathbb{Z}_2^{(1)}\rightarrow\mathbb{Z}$, namely, a lower level is included in the higher one.
\bibitem{note0} Note that  there exists only one possible spatial codimension $d_c=d-d_{FS}-1$ for $\mathbb{Z}$ topological charges,
namely, a chosen $S^{d_c}$ is demanded to enclose the FS from all its transverse dimensions. This is because $\mathbb{Z}$ topological charges do not require the anti-unitary symmetries, and an $S^{d'_c}$ with $d'_c<d_c$ can always be continuously shrunk to a gapped point by leaving the inversion symmetric plane in the $k$ space through the extra $d_c-d'_c$ dimensions, as shown in the right part of Fig.(\ref{Rotations}). Referring to the space-time codimension $p$ introduced in Ref.\cite{FS-classification, Volovik-book, Horava}, one can have $d_c=p-1$.
\bibitem{Note-FS}Local inversion-invariant dimensional FSs, such as a standard sphere centered at the origin of $k$ space, are regarded as points in the discussion of $\mathbb{Z}_2$ topological charges.
\bibitem{note2} While unitary spatial symmetries, such as rotation, inversion and reflection, ubiquitously exist in solid state systems, strictly speaking they are unavoidably broken by disorders. In contrast, anti-unitary symmetries like TRS and PHS may be preserved by certain types of disorders.
\bibitem{note1} It is noted that  the stability of a Fermi point with $|\nu|>1$ may be ensured by some unitary symmetries, such as rotation, inversion and reflection, for a condensed matter system, and the corresponding dispersion relation may also be non-linear, in contrast to the relativistic case.
\end{thebibliography}
\end{document}